\documentclass[runningheads,a4paper]{llncs}
\usepackage{amssymb}
\usepackage{lscape}
\usepackage{rotating}
\usepackage{graphicx}
\usepackage{comment}
\usepackage{url}
\newcommand{\keywords}[1]{\par\addvspace\baselineskip
\noindent\keywordname\enspace\ignorespaces#1}

\begin{document}

%\conferenceinfo{WXYZ '05}{date, City.} 
%\copyrightyear{2005} 
%
\title{Hello rootKitty:\\ A lightweight invariance-enforcing framework}
\titlerunning{Hello rootKitty: A lightweight invariance-enforcing framework}
\author{Francesco Gadaleta \and Nick Nikiforakis\and Yves Younan\and Wouter Joosen}
\authorrunning{F. Gadaleta et al.}
%\authorinfo{Francesco Gadaleta} {Katholieke Universiteit Leuven}{francesco.gadaleta@cs.kuleuven.be}
\institute{IBBT-DistriNet\\
Katholieke Universiteit Leuven\\
Celestijnenlaan 200A B3001 \\
Leuven, Belgium\\
\tt{francesco.gadaleta@cs.kuleuven.be}
%\mailsa
\\}

%\authorinfo{Nick Nikiforakis}{Katholieke Universiteit Leuven}{nick.nikiforakis@cs.kuleuven.be}
%\authorinfo{Yves Younan}{Katholieke Universiteit Leuven}{yvesy@cs.kuleuven.be}
%\authorinfo{Wouter Joosen}{Katholieke Universiteit Leuven}{wouter@cs.kuleuven.be}
%\authorinfo{Name2\and Name3}
%           {Affiliation2/3}
%           {Email2/3}

\maketitle

%%% ABSTRACT %%%%%%%%%%%%%%%%%%%%%%%%%%%%%%%
\begin{abstract}
In monolithic operating systems, the kernel is the piece of code that executes with the highest privileges and has control over all the software running on a host. A successful attack against an operating system's kernel means a total and complete compromise of the running system. These 
attacks usually end with the installation of a rootkit, a stealthy piece of software running with kernel privileges.
When a rootkit is present, no guarantees can be made about the correctness, privacy or isolation of the operating system.

In this paper we present \emph{Hello rootKitty}, an invariance-enforcing framework  which takes advantage of current virtualization technology to protect a guest operating system against rootkits. \emph{Hello rootKitty} uses the idea of invariance to detect maliciously modified kernel data structures and restore them to their original legitimate values. 
Our prototype has negligible performance and memory overhead while effectively protecting commodity operating systems from modern rootkits.

\end{abstract}

%%% INTRODUCTION %%%%%%%%%%%%%%%%%%%%%%%%%%%%%
\section{Introduction} \label{intro}
Operating systems consist of trusted software that executes directly on top of
a host's hardware providing abstraction, arbitration, isolation and security to the rest of the software. Due to their prominent position, operating systems have been
a common target of attackers who try to circumvent their protection mechanisms and modify them to their advantage. In the past, a program that allowed a user to elevate his access and become a system administrator (``root'') was called a \texttt{rootkit}. Today the meaning of rootkits has changed and is used to describe software that hides the attacker's presence from the legitimate system's administrator. 
Kernel-mode rootkits\footnote{Such rootkits appear very often in the form of device drivers (Microsoft Windows) or Loadable Kernel Modules (Linux kernel).} target the core of an operating system and thus they are the hardest to detect and remove.
In extreme cases, a kernel-mode rootkit may be introduced by a software bug in the kernel, triggered by a malicious or a benign but-exploitable process.
Regardless of the way the rootkit is introduced, the result is malicious code running with operating system privileges which can add 
and execute additional code or modify existent kernel code. The activities resulting from a successful attack can range 
from spamming and key-logging to stealing private user-data and disabling security software running on the host.
In the past, rootkits have also been used to turn their targets into nodes of a botnet as with Storm Worm~\cite{1387718} or to perform massive bank 
frauding~\cite{mcaffee}.

Even rootkits that do not introduce new code, but rather make use of existing fragments of code to fabricate their malicious functions, need to somehow have
these fragments executed in the order of their choosing~\cite{Hund2009}. Changing the control flow of the kernel involves either changing specific kernel objects such as function
pointers or overwriting existing fragments of code with new code.
Using the idea of modified kernel-data structures as a sign of rootkits, security researchers have developed several approaches to mitigate rootkits. Unfortunately, many of these are affected by considerable overhead~\cite{livewire,5} or miss a fundamental security requirement such as isolation\cite{hookscout}. 

Isolation is needed to prevent a countermeasure in the target system from being disabled/crippled by a potential attack.

Other countermeasures have been presented in which operating system kernels are
protected against rootkits by executing only authenticated (or validated) kernel
code \cite{SecVisor,NICKLE,5}. The aforementioned rootkit~\cite{Hund2009} that doesn't introduce new kernel code and re-uses fragments of authenticated code
bypasses such countermeasures.
In \cite{4} a countermeasure to detect changes of the kernel's control flow graph is presented; Anh et al. \cite{MAVMM} uses virtualization technology and emulation to perform malware analysis and \cite{HookSafe} protects kernel function pointers. 
Another interesting work is \cite{PoKer} which gives more attention to kernel rootkit profiling and reveals key aspects of the rootkit behavior by the analysis of compromised kernel objects. Determining which kernel objects are modiÞed by a rootkit not only provides an overview of the damage inflicted on the target but is also an important step to design and implement systems to detect and prevent rootkits.

A rising trend in security research is the use of virtualization technology for non-virtualization specific purposes~\cite{Gadaleta2009,QubesOS,Dewant2008,Criswell2007}. The property that makes virtualization particularly
attractive, from a security perspective, is that isolation is guaranteed by the current virtualization-enabled hardware. 
Using the appropriate instruction primitives
of such hardware, makes it straightforward to fully separate and isolate the target from the monitor system. 
In this paper we present \emph{Hello rootKitty}, a lightweight invariance-enforcing framework to mitigate kernel-mode rootkits in common operating system kernels. 
We start from the observation that many critical kernel data structures are
invariant. Many data structures used by rootkits to change the control-flow
of the kernel contain values that would normally stay unchanged for the lifetime
of a running kernel. Our protection system consists of a monitor that
checks the contents of data structures that need to be protected at regular times and detects whether their contents have changed. If a change is detected, our system 
warns the administrator of the exploited kernel and corrects the problem by restoring the modified data structures to their 
original contents. Our monitor runs inside a hypervisor and protects operating systems that are being virtualized. Due to the hardware-guarantees of
isolation that virtualization provides, an attacker has no way of disabling our monitor or tamper with the memory areas that our system uses.
\emph{Hello rootKitty} imposes negligible performance overhead on the virtualized system and it doesn't require kernel-wide changes other than 
a trusted module to communicate the invariant data structures from the guest operating system to the hypervisor. \emph{Hello rootKitty} can be integrated with
existing invariance-inferencing engines and protect commodity operating systems running on virtualized environments. Alternatively, our system can be used directly
by kernel developers to protect their invariant structures from rootkit attacks.

 The rest of the paper is structured as follows. Section~\ref{sec:problem} describes the problem of rootkits and presents our attacker model.
 Section~\ref{solution} presents our solution. In Section \ref{approach} we
present the architectural details of \emph{Hello rootKitty} followed by its implementation in Section~\ref{implementation}.
We evaluate our prototype implementation in Section~\ref{eval} and present its limitations in Section~\ref{evaluation}. We discuss related work on rootkit detection in Section~\ref{related} and conclude in Section~\ref{conclusion}.

%%% PROBLEM DESCRIPTION %%%%%%%%%%%%%%%%%%%%%%%%
\section{Problem description} \label{sec:problem}
In this section we describe common rootkit technology and we also present the model of the attacker that our system can detect and neutralize.

\subsection{Rootkits}

Rootkits are pieces of software that attackers deploy in order to hide their presence from a system. Rootkits can be classified according to the privilege-level 
which they require to operate. The two most common rootkit classes are: a) user-mode and b) kernel-mode.

User-mode rootkits run in the user-space of a system without the need of tampering with the kernel. In Windows, user-mode rootkits
commonly modify the loaded copies of the Dynamic Link Libraries (DLL) that each application loads in its address space~\cite{SymantecRootkits}. 
More specifically, an attacker can modify
function pointers of specific Windows APIs and execute their own code before and/or after the execution of the legitimate API call. In Linux, user-mode rootkits
hide themselves mainly by changing standard linux utilities, such as \texttt{ps} and \texttt{ls}. Depending on the privileges of the executing user, the rootkit
can either modify the default executables or modify the user's profile in a way that their executables will be called instead of the system ones 
(e.g.,by changing
the \texttt{PATH} variable in the Bash shell).

Kernel-mode rootkits run in the kernel-space of an operating system and are thus much stronger and much more capable. The downside, from an attacker's perspective,
is that the user must have enough privileges to introduce new code in the kernel-space of each operating system. In Windows, kernel-mode rootkits
are loaded as device-drivers and target locations such as the \texttt{call gate} for interrupt handling or the System Service Descriptor Table (SSDT).
The rootkits change these addresses (hooking) so that their code can execute before specific system calls. In Linux, rootkits can be loaded either
as a Loadable Kernel Module (LKM) or written directly in the memory of the kernel through the \texttt{/dev/mem} and \texttt{/dev/kmem} file~\cite{kmemRootkit}. These rootkits
target kernel-data structures in the same way that their Windows-counterparts do. Although this paper focuses on Linux kernel-mode rootkits, the concepts
introduced apply equally well to Windows kernel-mode rootkits.
An empirical observation is that kernel-mode rootkits need to corrupt specific kernel objects, in order to execute their own code and add malicious functionality 
to the victim kernel.
Studies of common rootkits \cite{packetstorm,4} show that most dangerous and insidious rootkits change function pointers in the system call table, interrupt descriptor table or in the file system, to point to malicious code. The attack is triggered by calling the relative system call from user space or by handling an exception or, in general, by calling the function whose function pointer has been compromised.
We report a list of rootkits which compromise the target kernel in Table~\ref{listrootkits}. 

\setlength{\textfloatsep}{0.5mm}
\begin{table}%[htdp]
\begin{center}
\scalebox{0.9}{
\begin{tabular}{|p{5cm}|p{5cm}|}
\hline
\bf{Rootkit} & \bf{Description} \\
\hline
\hline
Adore, afhrm, Rkit, Rial, kbd, All-root, THC, heroin, Synapsis, itf, kis   & Modify system call table \\ \hline
SuckIT & Modify interrupt handler \\ \hline
Adore-ng & Hijack function pointers of \texttt{fork(), write(), open(), close(), stat64(), lstat64()} and \texttt{getdents64()} \\ \hline
Knark & Add hooks to /proc file system \\
\hline
\end{tabular}
}

\end{center}
\caption{Hooking methods of common linux rootkits}
\label{listrootkits}
\end{table}
\

\subsection{Attacker Model}

In this work we first assume that the operating system which is being attacked is virtualized, i.e., it runs on top of a hypervisor
which has more privileges than the operating system itself. Virtualization guarantees isolation thus we assume that the guest
operating system cannot access the memory or code of the hypervisor. Our system detects the rootkit after it has been deployed, a fact which allows
our model to include all possible ways of introducing a rootkit in a system. Thus, a rootkit can be introduced either by: 
\begin{itemize}
\item A privileged user loading the rootkit as a Loadable Kernel Module
\item A privileged user loading the rootkit by directly overwriting memory parts through the \texttt{/dev/} memory interfaces
\item An unprivileged user exploiting a vulnerability in the kernel of the running operating system which will allow him to
execute arbitrary code
\end{itemize}
Finally, our system doesn't rely on secrecy so our model of the attacker
includes him being aware of the protection system.

%%% THE COUNTERMEASURE %%%%%%%%%%%%%%%%%%%%%%%
\section{Hello rootKitty: protecting kernel data against rootkits}
\label{solution}
In this section we describe our approach to detect rootkits that compromise function pointers or data structures residing in the kernel. 
%Our general approach is described in Section \ref{approach} and our implementation is discussed in Section \ref{implementation}. 

%%% APPROACH %%%%%%%%%%%%%%%%%%%%%%%%%%%%%%%
\subsection{Approach} \label{approach}
By studying the most common rootkits and their hooking techniques one can realize that they share at least one common characteristic. In order to achieve execution of their malicious code, rootkits overwrite locations in kernel memory which are used to dictate, at some point, the control-flow inside the kernel. 
Most of these locations are very specific (see Table~\ref{listrootkits}) and their values are normally invariant, i.e. they don't change over the normal execution of the kernel. Since these objects are normally invariant, any sign of variance can be used to detect the presence of rootkits. We use the terminology of ``critical kernel objects'' to name objects that can be used by an attacker to change the control-flow of the kernel. 
The approach of \emph{Hello rootKitty} is, given a list of invariant critical kernel objects, to periodically check them for signs of variance. When our countermeasure detects that the contents of an invariant critical kernel object have been modified, it will report an ongoing attack. Invariant critical kernel objects have been identified in several contributions, such as \cite{HookSafe,6,7,8}. The methods to detect invariance differ depending on the type of critical kernel object and are the following:

\begin{enumerate}
\item Static kernel objects at addresses hardcoded and not dependent on kernel compilation 
\item Static kernel objects dependent on kernel compilation (e.g., provided by \texttt{/boot/System.map} in a regular Linux kernel)
\item Dynamic kernel objects allocated on the heap by kmalloc, vmalloc and the rest of the kernel-specific memory allocation functions
\end{enumerate}

Identifying and protecting static kernel objects (type 1 and type 2) is
straightforward. During the installation of the operating system to be monitored, a virtual machine installer would know in advance whether the guest is of Windows or Unix type. This is the minimal information required to detect kernel objects whose addresses have been hardcoded (type 1). 
Moreover, the  Linux operating system provides \texttt{System.map}, where compilation-dependent addresses of critical kernel objects are stored (type 2). In contrast, identifying dynamic kernel objects (type 3) needs much more effort and depends on the invariance detection algorithm in place. Part of our countermeasure is a trusted module which operates in the guest operating system at boot time. Boot time is considered our root of trust. We are confident this to be a
realistic assumption. \footnote{Boot time ends right before calling \emph{kernel\_thread} which starts \emph{init}, the first userspace application of the Linux kernel. At this stage the kernel is booted, initialized and all the required device drivers have been loaded.}
From this point on, the system is considered to operate in an untrusted environment and a regular integrity checking of the protected objects is necessary to preserve the system's safety.
Given a list of invariant kernel objects, the trusted module communicates this data (virtual address and size) of the kernel objects to observe after boot, and stores them in the guest's address space. Then it will raise a hypercall in order to send the collected entries to the hypervisor. The hypervisor will checksum the contents mapped at the addresses provided by the trusted module and will store their hashes in its address space, which is not accessible to the guest. The trusted module is then forced to unload via a \emph{end-of-operation} message sent by the hypervisor. \emph{Hello rootKitty} doesn't accept objects after the kernel has booted, in order to prevent a possible Denial-Of-Service attack launched by
an attacker who is aware of the presence of our system. It is important to point out that \emph{Hello rootKitty} is not a invariance-detection system for critical kernel objects and thus it must be provided with a list of kernel objects on which it will enforce invariance. This list can be either generated by invariance detection systems~\cite{HookSafe,6,7,8} or manually compiled by kernel and kernel-module developers.

\begin{figure}[htbp] 
\begin{center}
\includegraphics[scale=0.5]{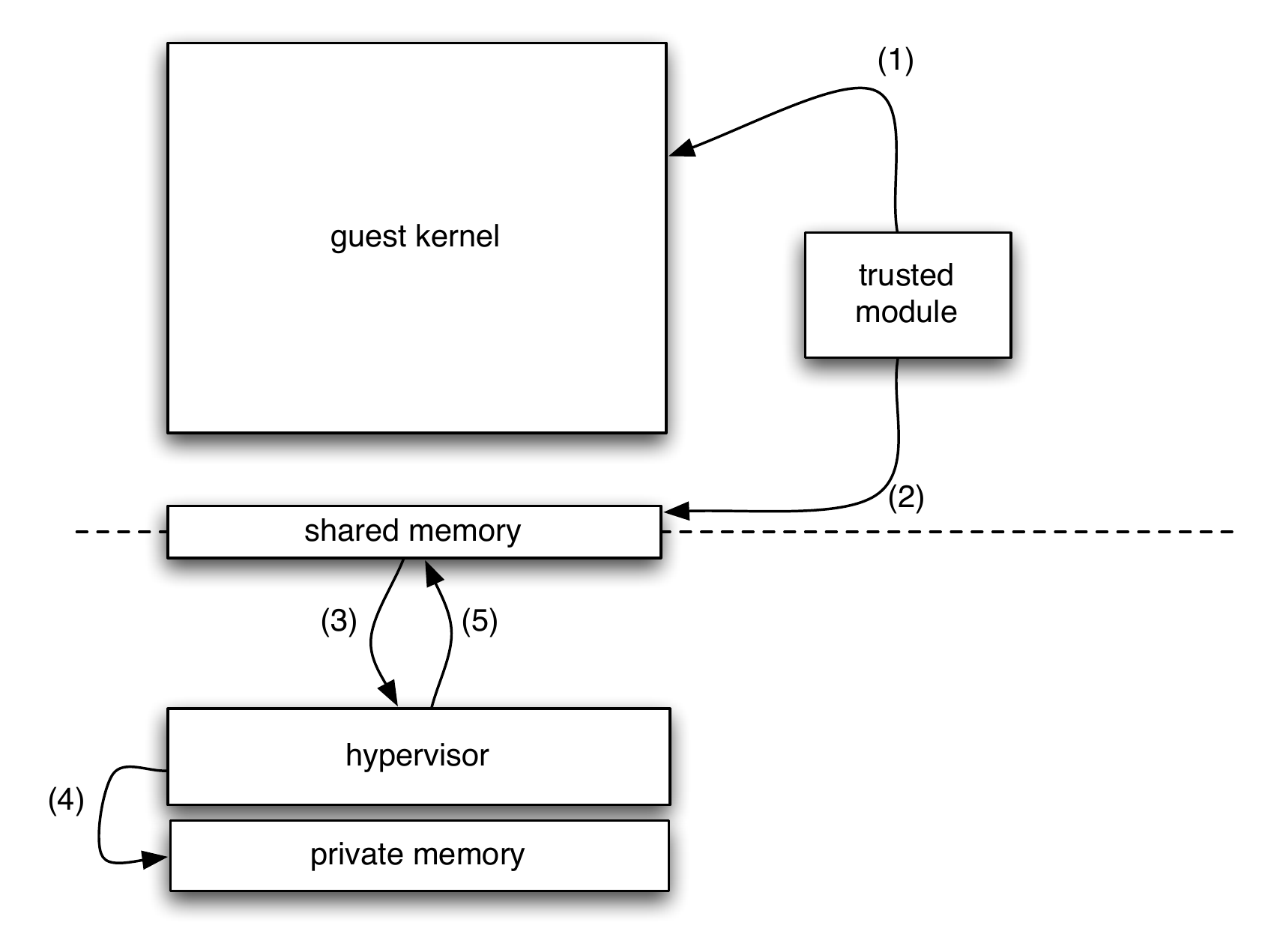}
\caption{{\bf High level view of trusted module-hypervisor interaction}}
\label{schema_tm_hyperv}
\end{center}
\end{figure}

Although implementing countermeasures in a separated virtual machine or within the hypervisor increases the degree of security via isolation, it often leads to higher performance overhead than the equivalent implementation in the target system. 
A challenging task is that of checking integrity outside of the target operating system while limiting the performance overhead. We achieve this by exploiting the regular interaction of a Virtual Machine Monitor and the guest operating system. In a virtualized environment the guest's software stack runs on a logical processor in VMX non-root operation~\cite{Intel2007}. This mode differs from the ordinary operation mode because certain instructions executed by the guest kernel may cause a \texttt{VMExit}. A \texttt{VMExit}, is a transition from VMX non-root mode to VMX root mode. After a \texttt{VMExit} the hypervisor will gain control of the CPU and will handle the exception. When the handler terminates the hypervisor performs a \texttt{VMEntry} and returns the control to the guest which will load the latest state of the logical processor and resume execution in VMX non-root mode.
Our countermeasure performs integrity checks every time the guest kernel writes to a control register (MOV\_CR* event) which in-turn causes a \texttt{VMExit}. 
Trapping this event is strategic because when virtual addressing is enabled, the upper 20 bits of control register 3 (CR3) become the page directory base register (PDBR). This register is fundamental to locate the page directory and the page tables for the current task. Whenever the guest kernel schedules a new process (process switch) the guest CR3 is modified. Performing integrity checks on the MOV\_CR* events is a convenient way to keep detection time and performance overhead to a minimum while guaranteeing a high level of security on protected objects. Moreover, this choice allows a constraint relaxation to improve performance even more by paying a small cost in terms of detection time. We provide more details for our constraint relaxation in Section \ref{implementation}.
Alternatively the hypervisor could check integrity randomly during the execution of the guest operating system. But this would not scale according to the guest system load as our current approach does.

%%% IMPLEMENTATION %%%%%%%%%%%%%%%%%%%%%%%%%%%
\subsection{Implementation}\label{implementation}
In this section we discuss the implementation details of \emph{Hello rootKitty}. 
We consider the choice of the hypervisor of critical importance in order to limit the overhead of the entire system. 
In fact, countermeasures implemented in virtualized environments are usually affected by considerable overhead which often prevents their deployment in actual
production systems.
We developed a prototype of our countermeasure in BitVisor, a tiny Type-I hypervisor~\cite{1508311} which exploits Intel VT and AMD-V instruction sets. Our target system runs a Linux kernel with version 2.6.35 and the trusted module has been implemented as a loadable kernel module for the Linux kernel. 
Our choice of BitVisor is mainly due to its memory address translation features. In BitVisor, the guest operating system and the Virtual Machine Monitor share the same physical address space. Thus, the VMM does not need any complex mechanism to provide translations from guest to host virtual addresses. The guest operating system will rely on the guest page table to perform translations from virtual to physical addresses. This considerably reduces the size of the hypervisor's code and has a very low impact on the overall performance. Unfortunately, in this specific architecture, the VMM can not directly use the guest page table. Translations of guest virtual addresses to host virtual addresses are thus performed by the cooperation of the trusted module and the hypervisor, as explained later in this section. 

Figure \ref{schema_tm_hyperv} presents a high-level view of \emph{Hello rootKitty}. Our system detects illegal modifications to invariant critical kernel objects 
in three phases which are described below.

\textbf{Communicating phase}
The trusted module executes in the guest's address space and communicates the addresses and sizes of critical kernel objects to be protected. 
In order to test and benchmark our system in a realistic way, we created an artificial list of critical kernel objects by allocating synthetic kernel data.
For each critical object the trusted module will retrieve its physical address by calling $\_\_pa(virtual\_address)$, a macro of the Linux kernel. 
If the kernel object is stored in one physical frame the trusted module will immediately collect the start address and the size. If the kernel object is stored on more than one physical frames the trusted module will store the relative list of physical addresses.
When the virtual addresses of all objects have been translated, a hypercall is raised which signals the hypervisor to start the integrity checking.

\textbf{Detection phase}
In order to detect changes the hypervisor needs to access the contents at the physical addresses collected by the trusted module. This is achieved by mapping the physical address and size of each object in its private memory and computing the signature of its actual contents. When all objects have been checksummed an end-of-operation 
flag is set in a memory area shared with the trusted module, which in turn will be unloaded. 
The checksum is performed by a procedure which implements MD5. This
cryptographic hash function provides the integrity guarantees needed for our
purposes. While stronger hash functions exist, we believe that the security and collision rate provided by MD5 are strong enough
to adequately protect our approach from mimicry attacks. 

\textbf{Repairing phase}
When the hypervisor detects that the signature of a protected object is different from the one computed the first time, two different behaviors are allowed: a) the system will report an ongoing attack or b) the system attempts to restore the contents of the compromised object if a copy has been provided by the trusted module. Since the hypervisor and the guest share the same physical address space, the hypervisor can restore the original content by mapping the physical address of the compromised object in its virtual space. The untampered value is then copied and control returns to the guest. The restoration of modified
critical data structures means that, while the rootkit's code is still present in the address space of the kernel, it is no longer reachable by the kernel control-flow and thus it is neutralized. Since switching from VMX-root to VMX-non-root causes a flush of the TLB, any code in the guest that was using the compromised object will perform the address translation
and memory load again and will thus load the restored value.
As previously mentioned, whenever task switching, the CR3 register's contents are changed. \emph{Hello rootKitty} traps
the MOV\_CR* event and checks the integrity of the critical kernel objects. This checking occurs outside of the guest operating system and thus can't be
influenced by it.
Since the number of kernel objects might be high, the hypervisor will perform the integrity checking of only a subset of objects. 
Control is then returned to the guest
kernel and another subset of critical kernel objects will be checked at the next MOV\_CR* event. While considerably improving the performance overhead, this relaxation obviously comes at a cost in terms of security and detection time. We do believe however, that the resulting detection ability of \emph{Hello rootKitty} remains
strong, a belief which we explore further in Section \ref{evaluation}.

%%%  EVALUATION %%%%%%%%%%%%%%%%%%%%%%%%%%%%%%
\section{Evaluation} \label{eval}
We implemented \emph{Hello rootKitty} in BitVisor (Ver 1.1) and the trusted module as a loadable kernel module of the Linux kernel. All experiments were performed on Intel Core 2 Duo 2 Ghz processor with 4GB of RAM. 

\subsection{Security Evaluation}

In order to evaluate whether \emph{Hello rootKitty} would detect a real rootkit we downloaded
and installed a minimal rootkit~\cite{kernel_hijack} which hijacks a system-call entry, specifically the \texttt{setuid} systemcall,
from the system-call table. When the \texttt{setuid} system call is invoked with the number \texttt{31337} as an argument, the 
rootkit locates the kernel structure for the calling process and elevates its permissions to ``root''. The way of
hijacking entries in the system-call table is very common among rootkits (see Table~\ref{listrootkits}) since it provides the rootkit a convenient
and reliable control of sensitive system calls.
 The critical kernel object that the rootkit modifies is the system-call table which normally remains invariant throughout the lifetime
of a specific kernel version. The Linux kernel developers have actually placed this table in read-only memory, however the rootkit
circumvents this by remapping the underlining physical memory to new virtual memory pages with write permissions.

 Before installing the rootkit, we gave as input to our trusted module, the address of the invariant system-call table
and its size. Since \emph{Hello rootKitty} is an invariance-enforcing framework and not an invariance-discovering system, the invariant critical kernel-objects and their size must
be provided to it from an external source. This source can either be automatic invariance-discovering systems or kernel programmers who wish
to protect their data structures from malicious modifications. Once our system was booted we loaded the rootkit in the running kernel. When the next \texttt{MOV} to control-registered occured, the system trapped
into the hypervisor and \emph{Hello rootKitty} detected the change on the invariant system-call table. After reporting the attack, the system repaired
the system-call table by restoring the system-call entry with the original memory address. This means, that while the rootkit's code is still loaded
in kernel-memory it is no longer reachable by any statement and thus inactive.

This shows, that \emph{Hello rootKitty} can detect rootkits and repair the kernel provided that a) the kernel objects used by the rootkit to achieve control
are invariant and b) the utilized kernel objects are included in the list of invariance that is given to our system.

\subsection{Performance benchmarks} \label{performance}
According to $slabtop$, a Linux utility which displays kernel slab cache
information, approximately 15,000 kernel objects are allocated during system's lifetime, 
75\% of which smaller than 128 bytes. These numbers are never exceeded in other
detection systems. Thus in
order to measure the overhead introduced by our countermeasure we instrumented
the trusted module to create 15,000 kernel objects each of 128 bytes and then
performed different types of benchmarks. In order to avoid checking all 15,000 objects
at each \texttt{VMExit} we check each time a different subset of the object set. This
parameter is configurable and its value depends on the priorities of each installation (performance versus
detection time). We measure real (wall-clock) timings
in a virtualized environment to compensate for the inaccuracy of time measurements 
within the virtual machine (i.e. the guest's timers are paused when the hypervisor is performing any other operation). 
We collected results from ApacheBench \cite{apachebench} sending requests on a local
webserver running lighttpd (Table \ref{apache}) and from SPECINT 2000 as macrobenchmarks to
estimate the delay perceived by the user (Table \ref{specbench}). 
Lastly, we collected accurate timings of microbenchmarks from lmbench (Table \ref{microbench}).
The macrobenchmarks show that our system imposes neglibible overhead on the SPEC applications (0.005\%)
allowing its widespread adoption as a security mechanism in virtualized systems.

%\begin{landscape}
%\begin{table*}[htdp]
\begin{sidewaystable}
\begin{center}
\begin{tabular}{|l|l|l|l|l|l|l|l|} \hline
\multicolumn{8}{|c|}{\bf{Processes - times in microseconds - smaller is better}} \\ \hline
                      & open clos  & slct TCP & sig inst  & sig hndl &  fork proc  &  exec proc  & sh proc \\\hline
no counterm.	   &16.6& 3.08& 0.48& 2.41& 1222& 4082 & 16.K \\
Hello rootKitty &16.5  & 3.09& 0.48& 2.47& 1724& 5547 & 18.K \\ 
\hline
\bf{overhead (\%)}    &0.6\% &   0.3\% &  0\%  &  2.5\%& 41.0\%&  35.8\%& 12.5\% \\ \hline

%\multicolumn{8}{|c|}{\bf{Context switching - times in microseconds - smaller is better}} \\ \hline
%                            & 2p/0K &  2p/16K & 2p/64K  & 8p/16K  &  8p/64K & 16p/16K  & 16p/64K \\ \hline 
%no counterm. 	   & 6.1900 & 6.0100 & 5.8500  &  25.4   &  25.5   &  26.1    &  25.6 \\
%hello rootKitty   & 8.6300 & 6.3900 & 7.3600  & 122.3   &  123.2  &  122.0   & 121.8 \\ 
%\hline
%\bf{overhead (\%)}      & 39.4\% &  6.3\% &  25.8\% &  388\% &   382\% &  367\%   & 375\%\\ \hline
\multicolumn{8}{|c|}{\bf{File and VM system latencies in microseconds - smaller
	is better}} \\ \hline
                     & 0K File create  & 0K File delete & 10K File create & 10K
					 File delete & Mmap latency & Prot fault &  Page fault
					 \\ \hline
no counterm.   	    & 26.0 &  21.5  &  99.9  & 28.2   &  62.2K & 4.355 & 9.32010  \\
Hello rootKitty	& 26.4 &   21.3 &  99.8  & 27.8   & 66.5K  & 4.444 & 9.84780  \\
\hline
\bf{overhead (\%)}  & 1.53\% & -0.93\%&  -0.1\%& -1.43\%& 6.9\%&  2.0\%&  5.5\%   \\ \hline
\multicolumn{8}{|c|}{\bf{Local Communication bandwidth in MB/s - bigger is better}} \\ \hline
	                         & TCP & File reread &  Mmap reread & Bcopy(libc) & Bcopy (hand) & Mem read  &  Mem write  \\ \hline
no counterm.      & 2401 & 313.0 & 4838.1 &  617.5  & 616.1 & 4836 &  698.7 \\
Hello rootkitty  & 2348 & 313.2 & 4885.0 & 619.7   & 618.8 & 4842 & 697.8  \\
\hline
\bf{overhead (\%)}     & 2.2\%& -0.06\%& -0.93\%&  -0.32\%&  -0.43\%&  -0.12\%& 0.12\%\\ \hline
\end{tabular}
\caption{Performance overhead of Hello rootKitty in action on lmbench benchmarks}
\label{microbench}

\vspace{3\baselineskip}
%\begin{table*}

%\begin{center}
\begin{tabular}{|l|l|l|l|}
\hline
\bf{Benchmark} & \bf{no counterm.} & \bf{Hello rootKitty} & \bf{Perf.overh.} \\
\hline
Time 				 &     7.153 (sec)   &    7.261 (sec)  &  1.50\% \\
\hline
Requests per second  &  13981.10 (num/sec) & 13771.43 (num/sec)  & 1.52\% \\
\hline
Time per request     & 3.576 (ms)        & 3.631 (ms)   &  1.54\% \\
\hline
Time per concurrent request  & 0.072 (ms)& 0.073 (ms)  & 1.4\% \\
\hline
Transfer rate        & 52534.36 (Kbytes/sec) & 51746.51 (Kbytes/sec)  & 1.52\%
\\
\hline
\end{tabular}
\end{center}
\caption{{\bf Results of ApacheBench sending 100000 requests, 50 concurrently on local lighttpd webserver}}
\label{apache}
%\end{table*}
\end{sidewaystable}

Microbenchmarks show a consistent overhead on process forking, as expected. 
In Table \ref{microbench} we do not report measurements of context switching
latencies because the numbers produced by this benchmark are inaccurate
\cite{lmbenchbug,lmbenchevil}. An improvement of local communication
bandwidth is due to the slower context switching which has the side effect of
slightly increasing the troughtput of file or mmap re-reading operations.  

%\begin{table*}[htdp]
%\begin{center}
%\begin{tabular}{|c|c|c|c|}
%\hline
%\bf{Benchmark} & \bf{no counterm.(ms)} & \bf{hello rootKitty!(ms)} & \bf{Perf. overh.} \\
%\hline
%compression   & 21.32  & 21.40  & 0.4\% \\
%\hline
%decompression & 6.73   &  7.33  & 8.9\%\\
%\hline
%compiling   & 394.3  & 421.2  & 7.0\% \\
%\hline
%\end{tabular}
%\end{center}
%\caption{{\bf Performance overhead of hello rootKitty! in action on Linux kernel
%	compression/decompression (bzip/bunzip) and compiling}}
%\label{kernel}
%\end{table*}

\begin{comment}
\begin{table*}%[htdp]
\addtolength{\belowcaptionskip}{-3mm}
\addtolength{\floatsep}{-3mm}
\begin{center}
\begin{tabular}{|l|l|l|l|}
\hline
\bf{Benchmark} & \bf{no counterm.(sec)} & \bf{Hello rootKitty(sec) } & \bf{Perf. overh.} \\
\hline
bh      & 159.84 &  160.58 & 0.46\% \\
\hline
bisort  & 157.22 & 157.85 & 0.40\%\\
\hline
health  & 13.58 & 13.60 &  0.14\% \\
\hline
mst     & 11.47 & 11.67 & 1.74\%  \\ 
\hline
perimeter & 4.13 & 4.15 & 0.48\% \\
\hline
treeadd & 35.11 &  35.30 & 0.54\% \\
\hline
tsp & 36.14 & 36.24 & 0.27\% \\
\hline
\end{tabular}
\end{center}
\caption{{\bf Olden benchmarks of Hello rootKitty in action }}
\label{oldenbench}
\end{table*}
\end{comment}

\begin{table*}%[htdp]
\addtolength{\belowcaptionskip}{-3mm}
\begin{center}
\begin{tabular}{|l|l|l|l|}
\hline
\bf{Benchmark} & \bf{no counterm.(sec)} & \bf{Hello rootKitty(sec) } & \bf{Perf. overh.} \\
\hline
164.gzip & 204 &  204 &  0\% \\
\hline
175.vpr  &  138 & 142 &  2.8\%\\
\hline
176.gcc  & 88.7 & 89.0 &  0.3\% \\
\hline
181.mcf   & 86.4 & 86.7 & 0.34\%  \\
\hline
197.parser & 206 & 207 & 0.5\% \\
\hline
256.bzip2  & 179  &  179 & 0\% \\
\hline
300.twolf  & 229 & 229 &  0\% \\
\hline
\textbf{Average} & 161.6 & 162.4 & 0.005\%\\
\hline
\end{tabular}
\end{center}
\caption{{\bf SPEC2000 benchmarks of Hello rootKitty in action }}
\label{specbench}
\end{table*}

%\begin{figure*}[htbp]
%\begin{center}
%\includegraphics[scale=0.5]{sec_eval}
%\caption{{\bf Schema of a possible attack by cooperative processes}}
%\label{sec_eval}
%\end{center}
%\end{figure*}

\subsection{Memory overhead} \label{memory}
Memory overhead is proportional to the number of protected objects. The
data structure needed to store information for integrity checking is 20 bytes
long (64-bit kernel object physical address, 32-bit kernel object size, 32-bit
checksum, 32-bit support flags used by the hypervisor) \footnote{In
order to repair the compromised object, the hypervisor needs to store the
object's original content too. This may increase the memory overhead.}. 
Protecting 15,000 objects costs 193KB when the original content is not
provided and 2168 KB otherwise.
Moreover, every time a subset of the list of objects is checked the hypervisor
needs to map each object from the guest physical space to its virtual space. 
In our proof of concept the hypervisor will map 100 objects of 128 bytes each
every time a MOV\_CR event is trapped. This has an additional cost of 13KB. Thus
the overall cost in terms of memory is approximately 206KB (2181KB if a copy of the
original content is provided). Since the regular hypervisor allocates
128MB at system startup, the memory overhead is 1.5\%. 
The trusted module needs the same amount of memory. But since it will free
previously allocated memory after raising the hypercall, that memory would be regularly 
used by the kernel. 
%Moreover guests are normally equipped with a higher amount of memory than the hypervisor.

\subsection{Detection Time}

Due to the relaxation of integrity checking introduced in the earlier sections, it is
possible that the modified critical kernel object will not be in the current subset that
\emph{Hello rootKitty} checks. The current prototype of our system checks a subset of 100
objects at every change of a Control Register. For the total of 15,000 objects, this means
that in the worst case scenario \emph{Hello rootKitty} will detect the malicious modification
149 process switches later than the moment it happened. We found out that in a 
normally loaded system, this corresponds to approximately 6 seconds of wall-clock time. We 
believe that this is an acceptable security trade-off for the performance benefits that
relaxation offers.

\section{Limitations}
 \label{evaluation}

In this section we describe the limitations and possible weak points of our countermeasure.
Since \emph{Hello rootKitty} checks the critical kernel objects for invariance at every change of a Control Register (CR*), an
attacker can possibly compromise the scheduler of the operating system and avoid task switching, thus avoiding changes of the CR3.
The problem with this attack is that it effectively freezes the system, since the control can't be returned back from the kernel to
the running applications. A rootkit's main goal is to hide itself from administrators thus any rootkit behaving this way
will a) reveal that there is something wrong in the kernel of the running operating system and b) will never be able to intercept system calls
of running processes. These facts suggest that while the attack is possible, it is not probable.

Another attack might occur because of the relaxation explained in Section \ref{implementation} which improves the performance overhead but comes with a cost in terms of detection time. 
Since at any MOV\_CR event the hypervisor will check the integrity of a subset
of objects, several malicious processes in the guest might compromise the kernel
and restore the original contents before the hypervisor performs the checking.
We consider such an attack hard to accomplish because, although the hypervisor performs integrity checking in a deterministic fashion, the attacker has no information about the position of the compromised object in the hypervisor's memory space. 
A possible mitigation for this kind of attack is the randomization of the sequence in which blocks are checked.

A third possible way to compromise the guest kernel would be by corrupting critical kernel objects whose values legitimately change during the kernel's lifetime. Such
objects are not invariant and thus can't be included in the list of objects that \emph{Hello rootKitty} checks since our system is unable to differentiate
legitimate from non-legitimate changes. The majority though of existing kernel-mode rootkits modify invariant data structures thus our system 
reduces considerably the rootkit attack surface and prevents most rootkits from performing a successful attack.

Lastly, \emph{Hello rootKitty} depends on invariance inference engines to provide an accurate list of invariant critical kernel objects. 
Thus, if the inference engine used doesn't provide all the invariant critical kernel objects (false negatives), \emph{Hello rootKitty}
will be unable to detect attacks that occur in the non-reported kernel objects.

%%% RELATED WORK %%%%%%%%%%%%%%%%%%%%%%%%%%%%
\section{Related work} \label{related}
A number of efforts exist on detecting and preventing kernel malware. 
In this section we explore related work that attempts to protect
a kernel using specialized hardware, virtualization, code integrity and
profiling.

\textbf{Hardware-based countermeasures} \label{hardware}
Copilot \cite{copilot} is a kernel integrity monitor which detects illegal
modiﬁcations to a host’s kernel by fetching physical memory pages where kernel
data and code is stored. Although the monitor has negligible overhead, it needs a separate PCI-card to fetch pages of the running kernel.
Gibraltar \cite{6} is a system to infer kernel data structure invariants by fetching snapshots of kernel memory in a way similar to Copilot, by using a PCI-card. The violation of the inferred invariants is reported as a potential attack. 
While detecting malicious behavior, those countermeasures are limited by the usage of special hardware. A wide deployment of such systems is harder to achieve. 

\textbf{Kernel code integrity} \label{integrity}
A countermeasure specifically designed to prevent the execution of unauthorized code is described in \cite{SecVisor}. The system comes in the form of a tiny hypervisor which protects legacy OSes and ensures that only validated code of the guest can execute in kernel mode. 
Another rootkit prevention system is NICKLE \cite{NICKLE}, which prevents
unauthorized kernel code execution via memory shadowing. The hypervisor
maintains a shadow physical memory to store authenticated guest kernel code. At
runtime it determines if the instruction fetch is for kernel or user mode. After
verifying the code it will route the instruction fetch accordingly to shadow or
standard memory. Kernel rootkit attacks would be detected and prevented since invalidated code would attempt to run in kernel mode.   

A recent attack to bypass countermeasures against code injection attacks,
such as the Non-Executable stack countermeasure, is Return Oriented Programming (ROP) \cite{geometry}. 
In ROP, the attacker, instead of injecting malicious code in the address space of a vulnerable process, crafts his malicious payload by combining fragments of existing code.
This method of attacking has been used to create return-oriented rootkits which
re-use fragments of authorized kernel code for malicious purposes
\cite{Hund2009}. Such rootkits can bypass countermeasures like the ones proposed by \cite{NICKLE,SecVisor}.\\
Yin et al. \cite{hookscout} protect kernel function pointers from being compromised by rootkits. The approach consists of an analysis and a detection subsystems. The analysis subsystem keeps track of function pointer propagation in kernel memory via a whole-system emulator and generates the policy for hook detection. The detection subsystem resides on the target machine and detects violations of the inferred policy. Although the described countermeasure is binary-centric and can generate a hook detection policy without modifying any guest source code, it can be disabled by a rootkit attack since the detection system resides in the target machine.
A countermeasure which protects kernel hooks dynamically allocated from the heap is described in \cite{HookSafe}. Since protection of kernel objects needs byte-level granularity, obviously finer than the page-level granularity provided by commodity hardware, this countermeasure relocates kernel hooks to be protected to a dedicated page and then exploits the regular page-level protection of the MMU. Although the overhead is negligible, rootkit attacks that compromise non-control data would not be prevented.

\textbf{Analysis and profiling systems} \label{profiling}
Malware analysis can reveal important information about the way a rootkit
compromises data structures or how private data are stolen, allowing researchers to understand rootkit's behavior and design effective countermeasures. 
A virtual machine monitor designed for malware analysis as described in
\cite{MAVMM} extracts features like memory pages, system calls, disk and network
accesses from the analyzed program running in the guest.
Wang et al. \cite{hookmap} is an analysis tool against persistent rootkits, which compromise kernel hooks for hiding purposes. Those kernel hooks are first identified by monitoring the kernel-side execution path of system utility programs (e.g. ps, ls) and then reported as potential targets.
Identification of kernel hooks and extraction of the hook implanting mechanisms
via dynamic analysis is proposed in \cite{hookfinder}. A whole-system emulator
is used rather than a virtual machine monitor. 
A rootkit proﬁler is described in \cite{PoKer}. A VMM is used to log the rootkit hooking behavior, to
monitor targeted kernel objects, to extract kernel rootkit code and infer the potential impact on user-level programs.

\emph{Hello rootKitty} can be easily integrated with the systems described above in
order to perform integrity checking and detect illegal changes to those kernel
objects collected by the analysis tools.

\section{Conclusion}\label{conclusion}
%In this paper we described the control-flow changes that rootkits must perform in order to achieve code execution and how changes on invariant data structures can be used to identify a rootkit's presence. 
In this paper we demonstrated how the guaranteed isolation between a hypervisor
and a guest operating system can be used to build a non-bypassable invariance-enforcing framework. 
We realized our idea by designing and implementing \emph{Hello rootKitty}, a lightweight countermeasure to mitigate
kernel-mode rootkits in common operating system kernels. Upon detection of a
change of an invariant kernel object, 
\emph{Hello rootKitty} alerts the administrator of the guest operating system and proceeds to repair the kernel by restoring the
data structure to its original values. Due to this change, the rootkit's injected code is no longer reachable by any statements
in the kernel and thus can no longer affect the running kernel's operations. The evaluation of our prototype showed that
\emph{Hello rootKitty} can detect control-flow changes used by modern rootkits with negligible performance and memory overhead, making it a viable countermeasure for protecting operating systems in virtualized environments.

\paragraph{Acknowledgements:}
This research is partially funded by the Interuniversity Attraction Poles Programme Belgian State, Belgian Science Policy, IBBT and the Research Fund K.U.Leuven.

\bibliographystyle{plain}
\bibliography{rootkitty}
%\bibliographystyle{eurosys}
%\bibliography{rootkitty}  % biblio.bib is the name of the Bibliography in this case
\end{document}